\documentclass[%
prl,%
preprint%
 ,secnumarabic%
,amssymb, amsmath,nobibnotes,aps]{revtex4}
\usepackage{epsfig}%
\usepackage{graphicx}%
\expandafter\ifx\csname package@font\endcsname\relax\else
 \expandafter\expandafter
 \expandafter\usepackage
 \expandafter\expandafter
 \expandafter{\csname package@font\endcsname}%
\fi

\begin{document}

%\markboth{Authors' Names}
%{Instructions for Typing Manuscripts (Paper's Title)}

%%%%%%%%%%%%%%%%%%%%% Publisher's Area please ignore %%%%%%%%%%%%%%
%\catchline{}{}{}{}{}
%%%%%%%%%%%%%%%%%%%%%%%%%%%%%%%%%%%%%%%%%%%%%%%%%%%%%%%%%%%%%%%%%%%
\title{Very Special Relativity is incompatible with Thomas precession}%

\author{\footnotesize Suratna Das \footnote{suratna@prl.res.in} } 

\affiliation{Physical Research Laboratory, Ahmedabad 380009, India.}
\author{\footnotesize Subhendra Mohanty \footnote{mohanty@prl.res.in} } 

\affiliation{Physical Research Laboratory, Ahmedabad 380009, India.}

%\date{August 2001}%

%\maketitle

\begin{abstract}
Glashow and Cohen make the interesting observation that certain proper
subgroups of the Lorentz group like $HOM(2)$ or $SIM(2)$ can explain
many results of special relativity like time dilation, relativistic
velocity addition and a maximal isotropic speed of light. We show here
that such $SIM(2)$ and $HOM(2)$ based VSR theories predict an
incorrect value for the Thomas precession and are therefore ruled out
by observations. In VSR theories the spin-orbital coupling in atoms
turn out to be too large by a factor of 2. The Thomas-BMT equation
derived from VSR predicts a precession of electrons and muons in
storage rings which is too large by a factor of $10^3$. VSR theories
are therefore ruled out by observations.
\\
\end{abstract}

%\ccode{03.30.+p}
\maketitle
%%%%%%%%%%%%%%%%%%%%%%%%%%%%%%%%%%%%%%%%%%%%%%
\section{Introduction}
Glashow and Cohen \cite{vsr} have made the interesting observation
that certain proper subgroups of the Lorentz group like $HOM(2)$ or
$SIM(2)$ can explain many results of Special Relativity (SR) like time
dilation, relativistic velocity addition and a maximal isotropic speed
of light. Glashow and Cohen further suggest that particle physics
models can be constructed with only $HOM(2)$ or $SIM(2)$ invariance
(which they call VSR (Very Special Relativity)) which will give the
same dynamics as the full Lorentz invariant theory. These subgroups
($HOM(2)$ or $SIM(2)$) of Lorentz group along with either $P$, $CP$ or
$T$ generate the full Lorentz group \cite{vsr}. An interesting
application of this idea is to construct a $SIM(2)$ invariant mass
term for the neutrino using only the standard model left handed
neutrinos \cite{glashow}.  There are many theories that have been
constructed based on the $HOM(2)$ or $SIM(2)$ algebra. In \cite{NC} it
has been shown that Quantum Field Theories based on non-commutative
space-times provide a setting for non-trivial realizations of the VSR
algebras. VSR transformations can be generalized to curved space-time
\cite{Gibbons,Alvarez,Muck}. Super-symmetric theories based on SIM(2)
algebra have also been constructed \cite{SUSY1,SUSY2,SUSY3}.

One interesting point to note is that VSR theories differ from other
Lorentz violating theories where the Lorentz violation is governed by
a small parameter and as this parameter approaches to zero, one
regains the full Lorentz invariant theory. In VSR theories the Lorentz
violation is obtained by replacing the Lorentz group by its proper
subgroups.

Though the VSR theory has certain interesting consequences as
mentioned above, it is worth noticing at this point that it will be
incorrect to expect that a proper subgroup of the Lorentz group to
reproduce all the results of the full Lorentz group. For an example
consider the two $SU(2)$ subgroups generated by
$N_i=\frac12(J_i+iK_i)$ and $\tilde{N}_i=\frac12(J_i-iK_i)$ of the
Lorentz group. These two subgroups transform into each-other under
parity. Any one of the subgroup augmented with parity can generate the
full Lorentz group. However even for parity conserving processes it
would be naive to expect all the results of the full Lorentz group
$SO(3,1)\sim SU(2)\otimes SU(2)$ by taking only one of the $SU(2)$
subgroups generated by either $N_i$ or $\tilde{N}_i$. The invariance
of the Electromagnetic interaction involves the generators of both
$N_i$ and $\tilde{N}_i$ i.e. the full Lorentz group. Even for CP
conserving processes one can not expect only one of these subgroups
giving the same results as the full Lorentz group.

The above argument is also applicable for the subgroups ($HOM(2)$ and
$SIM(2)$) of the Lorentz group with which the VSR theories are
constructed. So any result of which can be derived using the
symmetries of the full Lorentz group need not necessarily follow by
using the transformations of the VSR subgroups. It is a remarkable
observation of Glashow and Cohen that many of the results of special
relativity like time dilation, velocity addition and the constancy of
the speed of light can be derived using only the $HOM(2)$ and $SIM(2)$
transformations and do not require the full $SO(3,1)$ group. We have
shown in this paper that this does not hold for another classic result
of Special Relativity namely Thomas precession.

We have arranged our paper in the following way : In Sec.~(\ref{vsr})
we verify the results of \cite{vsr} that VSR theories can mimic the
action of Lorentz transformations in boosts and in relativistic
velocity addition, though the VSR transformation parameters
required for velocity addition are not given either in \cite{vsr} or
in any follow up papers. In Sec.~(\ref{tho}) we find however that one
classic result of SR namely Thomas precession \cite{Thomas} which is
tested in the spin-orbit interaction of atoms does not come out
correctly in VSR.  We show here that such $SIM(2)$ and $HOM(2)$ based
VSR theories predict incorrect Thomas precession and are therefore
ruled out by observations of the fine splitting of atomic spectra. A
test of Thomas precession in a macroscopic setting is in the
spin-precession in external magnetic field which is described by the
Bargmann-Michel-Telegedi (BMT) equation \cite{BMT}. We show in
Sec.~(\ref{bmt}) that the VSR based theories lead to large corrections
in the BMT equation and are ruled out by the observations of
spin-precession in accelerators.

%---------------------------------------------------
\section{Very Special Relativity}
\label{vsr}

The Cohen-Glashow Very Special Relativity (VSR) \cite{vsr} is defined
as symmetry under certain proper subgroups of Lorentz group. The
minimal version of the VSR algebra contains, the subgroup $T(2)$ of
the Lorentz group, which is generated by $T_1 = K_x - J_y$ and $T_2 =
K_y+J_x$, where $J_i$ and $K_i$ $(i = x, y, z)$ are respectively
generators of rotations and boosts. $T(2)$ is an Abelian subalgebra of
Lorentz algebra $SO(1, 3)$ and can be identified with the translation
group on a two dimensional plane. The other larger versions of VSR are
obtained by adding one or two Lorentz generators to $T(2)$, which have
geometric realizations on the two dimensional plane. $E(2)$, the
3-parametric group of two dimensional Euclidean motion, generated by
$T_1$, $T_2$ and $J_z$, with the structure $[T_1, T_2] = 0$, $[J_z,
  T_1]= T_2$, $[J_z, T_2]=-T_1$. $HOM(2)$, the group of
orientation-preserving similarity transformations, or homotheties,
generated by $T_1$, $T_2$ and $K_z$, with the structure $[T_1, T_2] =
0$, $[T_1,K_z]=-T_1$, $[T_2,K_z] = -T_2$. $SIM(2)$, the group
isomorphic to the four-parametric similitude group, generated by
$T_1$, $T_2$, $J_z$ and $K_z$. The explicit forms of the VSR
generators are 
\begin{eqnarray}
T_1=\left(
\begin{array}{ccccccc}
0 & & 1 & & 0 & & 0\\
1 & & 0 & & 0 & & -1 \\
0 & & 0 & & 0 & & 0\\
0 & & 1 & & 0 & & 0
\end{array}
\right),\quad\quad
T_2=\left(
\begin{array}{ccccccc}
0 & & 0 & & 1 & & 0\\
0 & & 0 & & 0 & & 0 \\
1 & & 0 & & 0 & & -1\\
0 & & 0 & & 1 & & 0
\end{array}
\right),\nonumber\\
J_Z=\left(
\begin{array}{ccccccc}
0 & & 0 & & 0 & & 0\\
0 & & 1 & & 0 & & 0 \\
0 & & 0 & & -1 & & 0\\
0 & & 0 & & 0 & & 0
\end{array}
\right),\quad\quad
K_Z=\left(
\begin{array}{ccccccc}
0 & & 0 & & 0 & & 1\\
0 & & 0 & & 0 & & 0 \\
0 & & 0 & & 0 & & 0\\
1 & & 0 & & 0 & & 0
\end{array}
\right).
\label{generators}
\end{eqnarray}

%The $HOM(2)$ subgroup of the Lorentz group consists of 3 generators
%$T_1=K_x-J_y$, $T_2=K_y+J_x$ and $K_z$ where $K_i$'s and $J_i$'s are
%the generators of Lorentz boosts and 3-space rotations
%respectively. The $HOM(2)$ generators obey the commutation relations
%$\left[T_1,T_2\right]=0, \left[T_1,K_z\right]=-T_1,$ and
%$\left[T_2,K_z\right]=-T_2.$

The surprising result of VSR is that all the classic tests of Special
Relativity like Michelson-Morley experiment, time dilation, constant
isotropic maximal speed of light and velocity addition do not require
the full Lorentz group but can be derived using the generators of just
the $SIM(2)$ or $HOM(2)$ subgroups.

In VSR (Very Special Relativity) \cite{vsr} one can transform a
particle velocity in its rest frame $u_0=(1,0,0,0)$ to a moving frame
with four velocity $u=(\gamma_u,\gamma_u u_x,\gamma_u u_y,\gamma_u
u_z)$, where $\gamma_u=\frac{1}{\sqrt{1-{\mathbf u}^2}}$, by
transformations of the $HOM(2)$ group
\begin{equation}
L(u)u_0=u,
\label{lu0}
\end{equation}
where
\begin{equation}
L(u)=e^{\alpha T_1}e^{\beta T_2}e^{\phi K_z}
\label{lu}
\end{equation}
and the parameters are given by \cite{vsr}
\begin{eqnarray}
\alpha&=&\frac{u_x}{1-u_z},\nonumber\\
\beta&=&\frac{u_y}{1-u_z},\nonumber \\
\phi&=&
-\ln[\gamma_u(1-u_z)].
\label{parameters}
\end{eqnarray}

We can also define the $HOM(2)$ transformation matrices in terms of
the generators as
\begin{equation}
L(u)=e^{i \alpha T_1}e^{i \beta T_2}e^{i \phi K_z}
\label{lui}
\end{equation}
in that case one would have to include an extra factor of $-i$ with
the generators given in Eq.~(\ref{generators}).  The parameters of the
transformation remain the same as in Eq.~(\ref{parameters}) and are
always real.

These parameters are chosen in \cite{vsr} to give the same result as
the Lorentz transformation in SR. The $HOM(2)$ transformation
matrices are explicitly
\begin{eqnarray}
e^{\alpha T_1}=\left(
\begin{array}{ccccccc}
1+\frac{\alpha^2}{2} & & \alpha & & 0 & & -\frac{\alpha^2}{2}\\
\alpha & & 1 & & 0 & & -\alpha \\
0 & & 0 & & 1 & & 0\\
\frac{\alpha^2}{2} & & \alpha & & 0 & & 1-\frac{\alpha^2}{2}
\end{array}
\right),
\end{eqnarray}
\begin{eqnarray}
e^{\alpha T_2}=\left(
\begin{array}{ccccccc}
1+\frac{\beta^2}{2} & & 0 & & \beta & & -\frac{\beta^2}{2}\\
0 & & 1 & & 0 & & 0 \\
\beta & & 0 & & 1 & & -\beta\\
\frac{\beta^2}{2} & & 0 & & \beta & & 1-\frac{\beta^2}{2}
\end{array}
\right),
\end{eqnarray}
\begin{eqnarray}
e^{\phi K_z}=\left(
\begin{array}{ccccccc}
\cosh\phi & & 0 & & 0 & & \sinh\phi\\
0 & & 1 & & 0 & & 0 \\
0 & & 0 & & 1 & & 0\\
\sinh\phi & & 0 & & 0 & & \cosh\phi
\end{array}
\right),
\end{eqnarray}
which altogether yield the VSR transformation given in Eq.~(\ref{lu})
as
\begin{eqnarray}
L(u)=\left(
\begin{array}{ccccccc}
\gamma_u & & \frac{u_x}{1-u_z} & &  \frac{u_y}{1-u_z} & & \gamma_u\frac{u_z-u^2}{1-u_z}\\
\gamma_u u_x & & 1 & & 0 & & -\gamma_u u_x\\
\gamma_u u_y & & 0 & & 1 & & -\gamma_u u_y\\
\gamma_u u_z & & \frac{u_x}{1-u_z} & &  \frac{u_y}{1-u_z} & &  \gamma_u\frac{1-u^2}{1-u_z}
\end{array}
\right).
\end{eqnarray}
It is to be noted here that by construction this transformation is
quite different from the well known Lorentz boost given by 
\begin{eqnarray}
\Lambda(u)=\left(
\begin{array}{ccccccc}
\gamma_u & & \gamma_uu_x & &  \gamma_uu_y & & \gamma_uu_z\\
\gamma_u u_x & & 1+\frac{\left(\gamma_u-1\right)u_x^2}{u^2} & & \frac{\left(\gamma_u-1\right)u_xu_y}{u^2} & &\frac{\left(\gamma_u-1\right)u_xu_z}{u^2} \\
\gamma_u u_y & & \frac{\left(\gamma_u-1\right)u_xu_y}{u^2} & & 1+\frac{\left(\gamma_u-1\right)u_y^2}{u^2} & &\frac{\left(\gamma_u-1\right)u_yu_z}{u^2} \\
\gamma_u u_z & & \frac{\left(\gamma_u-1\right)u_xu_z}{u^2} & & \frac{\left(\gamma_u-1\right)u_yu_z}{u^2} & & 1+\frac{\left(\gamma_u-1\right)u_z^2}{u^2}
\end{array}
\right).
\end{eqnarray}

 The inverse transformation $L^{-1}(u)=e^{-\phi K_z}e^{-\beta
  T_2}e^{-\alpha T_1}$ takes a particle from a moving frame to its
  rest frame.
%----------------------------------------------
\subsection{Velocity addition in VSR}

Velocity addition law of Special Relativity (SR) is crucial in
ensuring that the speed of light is maximal and same in all inertial
reference frames. Hence to have an alternative theory of SR, the
new theory should also produce the same result for velocity addition.

Suppose a particle is moving with velocity $\mathbf{u}$ in an
inertial frame $(S^{\prime})$ which is moving with a velocity
$\mathbf{v}$ with respect to another inertial frame $(S)$. According
to SR applying two successive boosts $\Lambda(v)\Lambda(u)$ on the
rest frame of the particle gives the velocity addition law.

In VSR the velocity addition can not be given by successive $HOM(2)$
transformations $L(v)L(u)$ like that of in SR because the form of the
VSR transformation operator depends upon the reference frame unlike
Lorentz transformation of SR. So the operator which boosts a particle
at rest with velocity $\mathbf{u}$ given in Eq.~(\ref{lu}) is not the
same as the operator $L(v,u)$ which transforms a particle with
velocity $\mathbf{u}$ by a boost parameter $\mathbf{v}$. Such a
transformation can be constructed in $HOM(2)$
\begin{equation}
L(v,u)u=w
\end{equation}
(where $w$ is the relativistic sum of $u$ and $v$) with the general properties
\begin{equation}
L(v,0)= L(v)
\end{equation}
and
\begin{equation}
L(0,u)=I.
\end{equation}
For example the $HOM(2)$ transformation which boosts a particle with
velocity ${\bf u}=(u_x,u_y,u_z)$ by the boost parameter
${\mathbf v}=(v_x,0,0)$ is
\begin{equation}
L(v,u)=e^{\alpha^\prime T_1}e^{\beta^\prime T_2}e^{\phi^\prime K_z}
\end{equation}
with parameters $\alpha^\prime, \beta^\prime, \phi^\prime$ chosen as,
\begin{eqnarray}
\alpha^\prime&=&\frac{u_x-\left(u_x+v_x\right)\gamma_v}{u_z+\left(1+u_xv_x\right)\gamma_v}, \nonumber\\
\beta^\prime&=&0 ,\nonumber\\
\phi^\prime &=&-\frac{1-u_z}{u_z+(1+u_xv_x)\gamma_v}.
\end{eqnarray}
One can check explicitly that
\begin{eqnarray}
L(v,u)u&=&\left(\gamma_u\gamma_v(1+u_xv_x),\,\,\gamma_u\gamma_v(u_x+v_x),\,\,\gamma_uu_y,\,\,\gamma_uu_z\right)^{\rm T},
\end{eqnarray}
which is the correct relativistic velocity addition result.

To get the relativistic velocity addition when ${\mathbf v}=(0,v_y,0)$
the parameters will be
\begin{eqnarray}
\alpha^\prime&=&0\nonumber \\
\beta^\prime&=&\frac{u_y-\left(u_y+v_y\right)\gamma_v}{u_z+\left(1+u_yv_y\right)\gamma_v}, \nonumber\\
\phi^\prime &=&-\frac{1-u_z}{u_z+(1+u_yv_y)\gamma_v}.
\end{eqnarray}

However if ${\mathbf v}=(0,0,v_z)$ i.e. $S^\prime$ is moving in the
positive $z$ direction with respect to $S$ frame, then
$L(v_z,u)=L(v_z)$ i.e. $\alpha^\prime=0$, $\beta^\prime=0$ and
$\phi^\prime=\phi$ and in this case $L(v_z)u$ will give the correct
relativistic addition of velocities.

The transformation parameters
$\alpha^\prime,\,\,\beta^\prime,\,\,\phi^\prime$ that boosts a particle
with velocity ${\mathbf u}=(u_x,u_y,u_z)$ to ${\mathbf
  v}=(v_x,v_y,v_z)$ are algebraically complicated.
%-----------------------------------------------
\section{Thomas precession and spin-orbital coupling in VSR} 
\label{tho}

Thomas precession is a result of the property of Lorentz
transformation that two successive Lorentz boosts along different
directions can be combined as a single Lorentz boost and a
rotation. This extra rotation experienced by an accelerating particle
with non-zero spin is interpreted as due to an effective spin-orbit
coupling which changes the energy levels of quantum states and causes
extra precession in classical accelerating spinning bodies. A brief
derivation of Thomas precession is given in Appendix~(\ref{thomas-SR})
following \cite{jackson}. To derive Thomas precession we need to
calculate in SR
\begin{eqnarray}
A_T^{\rm SR}v(t)\equiv\Lambda({\mathbf v}+{\mathbf \delta {\mathbf v}})\Lambda^{-1}({\mathbf
  v})v(t)=(I-\Delta {\mathbf v}\cdot {\mathbf K}-\Delta {\mathbf \Omega}\cdot {\mathbf J})v(t),
\end{eqnarray}
which is discussed in Eq.~(\ref{thomas_sr}). Here $v(t)$ is the
particle's velocity at space-time position $x_0(t)$ at time $t$ and
$\Delta {\mathbf \Omega}$ is interpreted as Thomas precession. Also
${\mathbf K}\equiv (K_x, K_y, K_z)$ and ${\mathbf J}\equiv (J_x, J_y,
J_z)$ are the Lorentz boosts and rotations respectively.

Following the same argument, to determine Thomas precession in VSR the
required transformation will be
\begin{eqnarray}
L(v+\delta v)L^{-1}(v)v(t).
\end{eqnarray} 
Now since we have already chosen parameters of $L(u)$ to satisfy
Eq.~(\ref{lu0}), we no longer have any more freedom in the choice of
parameters. Hence we require to calculate the following transformation

%Now in VSR theory if we connect two velocities at different times with
%VSR transformations we have
%\begin{equation}
%v(t+\delta t)=A^{\rm VSR}_T v(t),
%\end{equation}
%and here we have
\begin{equation}
A^{\rm VSR}_T=L(v+\delta v)L^{-1}(v),
\end{equation}
where $L(v+\delta v)v_0=v(t+\delta t)$ and $L(v)v_0=v(t)$ are the VSR
transformation matrices which take the electron from its rest frame to
the rest frame of the nucleus at times $t+\delta t$ and $t$
respectively. Using the form of $L(v)$ given in Eq.~(\ref{lu}) we can
calculate $A^{\rm VSR}_T$ in first order of $\delta v_i$ which turns
out to have the form
\begin{eqnarray}
A_T^{\rm VSR}&=&\left(
\begin{array}{ccccccc}
1 & & \gamma^2\delta v_x & & \delta v_y & & -\gamma^2v_x\delta v_x\\
\gamma^2\delta v_x & & 1 & & 0 & & -\gamma^2\delta v_x \\
\delta v_y & & 0 & & 1 & & -\delta v_y \\
-\gamma^2v_x\delta v_x & & \gamma^2\delta v_x & & \delta v_y & & 1
\end{array}
\right)\nonumber\\
&=&I-\Delta\mathbf{v}_{\rm VSR}\cdot{\mathbf K}-\Delta{\mathbf \Omega}_{\rm VSR}\cdot {\mathbf J},
%&=&I+\left( \gamma^2\delta v_x,\delta v_y,-\gamma^2v_x\delta v_x\right)\cdot\mathbf{K}-\left(-\delta v_y,\gamma^2\delta v_x,0\right)\cdot\mathbf{J}.
\end{eqnarray}
where $\Delta\mathbf{v}_{\rm VSR}=-\left( \gamma^2\delta v_x,\delta
v_y,-\gamma^2v_x\delta v_x\right)$ and $\Delta{\mathbf \Omega}_{\rm
VSR}=\left(-\delta v_y,\gamma^2\delta v_x,0\right)$. Following
Eq.~(\ref{thomas-freq}) the angular velocity of the electron will be
\begin{equation}
\mbox{\boldmath$\omega$}_{\rm VSR}=-\frac{{\Delta \mathbf{\Omega}_{\rm VSR}}}{\delta t}=\left(a_y,-\gamma^2 a_x, 0\right).
\label{thomas-freq-vsr}
\end{equation}
In a circular orbit the acceleration is always radial, hence
$a_x=0$ for instantaneous velocity in $x$ direction. Therefore
the precession frequency in VSR turns out to be
\begin{equation}
\mbox{\boldmath$\omega$}_{\rm VSR}= \left(a_y,0,0\right),
\label{vsr-freq}
\end{equation}
It can be seen from the above equation that in this case there are
rotations around $x$ axis but no rotation around $z$ axis. The
spin-orbit coupling term due to this VSR precession is
\begin{eqnarray}
U_{\rm VSR}&=&\mbox{\boldmath$\omega$}_{\rm VSR}\cdot {\bf s}= s_x a_y\nonumber\\
&=&s_x  y\frac{1}{mr}\frac{dV}{dr}.
\end{eqnarray}
Hence the total spin interaction energy in case of VSR would be
\begin{eqnarray}
H_{SO}^{\rm VSR}={\frac{g}{2 m^2} } {\bf s}\cdot {\bf L}
 \frac{1}{ r } \frac{dV}{dr}+s_x  y\frac{1}{mr}\frac{dV}{dr}.
\end{eqnarray}
If the electron has a spin state where $\langle s_x \rangle=0$ then
the contribution to the spin-orbital energy of the electron will come
only from Eq.~(\ref{sob}) and will turn out to be too large by a
factor of 2 compared with the experimental results.
%---------------------------------------------------
\section{Thomas-BMT equation of spin precession in SR and VSR}
\label{bmt}

In a macroscopic setting, such as particle accelerators, Thomas
precession plays a significant role in the precession of particles
circulating in an external magnetic field. The total precession
frequency is from a combination of Larmor frequency due to the
particles magnetic moment and the Thomas precession due to the
acceleration involved in the circular motion. The BMT equation
\cite{BMT} which governs the spin precession in an external field is
tested in accelerators where the precession rate of particles is
measured to determine their anomalous magnetic moments. In this
section we show that the Thomas precession in VSR theories derived
earlier also modifies the BMT equation and leads to a precession
frequency of particles which is not observed.

 The equation of motion in an external magnetic field ${\mathbf B}$,
 in the lab-frame can be given by Eq.~(\ref{ds-dt}), where now we have
\begin{eqnarray}
\left.\frac{d {\bf s}}{dt}\right|_{\rm lab-frame}&=& \left.\frac{1}{\gamma}\frac{d {\bf s}}{d\tau}\right|_{\rm e-frame} +
\mbox{\boldmath$\omega$}_{T}\times {\bf s}\nonumber \\
&=&\left(\mbox{\boldmath$\omega$}_{L}+\mbox{\boldmath$\omega$}_{T}\right)\times{\bf s},
\end{eqnarray}
where $\tau$ is the proper time in particle's rest frame and
$\mbox{\boldmath$\omega$}_{L}\equiv-\frac{ge}{2m\gamma}{\mathbf
  B^{\prime}}$ and ${\mathbf B^{\prime}}\equiv\gamma{\mathbf
  B}_\perp+{\mathbf B}_\parallel$ is the effective magnetic field
realized by the particle in its rest frame and the parallel and
perpendicular components are with respect to the instantaneous
velocity of the particle. For simplicity we take the applied external
magnetic field in the $z$ direction and the instantaneous velocity in
the $x$ direction as considered while discussing Thomas precession and
therefore we get
\begin{eqnarray}
\mbox{\boldmath$\omega$}_{L}=-\frac{ge}{2m}B_z\widehat{{\mathbf k}}.
\end{eqnarray}
The frequency arising due to Thomas precession
$(\mbox{\boldmath$\omega$}_{T})$ can be obtained from
Eq.~(\ref{thomas-freq}). A particle moving in circular orbit in $x-y$
plane under the influence of an external magnetic field will have an
acceleration
%now have the form
%\begin{eqnarray}
%\mbox{\boldmath$\omega$}_{T}&=&\frac{\gamma^2}{\gamma+1}({\mathbf a}\times{\mathbf v})\nonumber \\
%&=&\left(0,\frac{\gamma^2}{\gamma+1}a_zv_x,-\frac{\gamma^2}{\gamma+1}a_yv_x\right)
%\label{thomas-freq1}
%\end{eqnarray}
%where the instantaneous velocity has been taken in the $x$ direction
%and the acceleration of the particle is due to the applied magnetic
%field in the accelerator which can be derived as
\begin{eqnarray}
{\mathbf a}=\frac{e}{m\gamma}({\mathbf v}\times{\mathbf B})=-\frac{e}{m\gamma}v_xB_z\,\widehat{{\mathbf j}},
\label{accelr}
\end{eqnarray}
%Hence the frequency for the Thomas precession given in
%Eq.~(\ref{thomas-freq}) will be
which yields
\begin{eqnarray}
\mbox{\boldmath$\omega$}_{T}%&=&\left(\frac{\gamma^2}{\gamma+1}\right)\frac{e}{m\gamma}\left[({\mathbf v}\times{\mathbf B})\times{\mathbf v}\right]\nonumber \\
=(\gamma-1)\frac{e}{m\gamma}B_z\widehat{{\mathbf k}}.
%&=&(\gamma-1)\frac{e}{m\gamma}{\mathbf B_\perp}.
\end{eqnarray}
Therefore the total precession of the charged particle
is
\begin{eqnarray}
\mbox{\boldmath$\omega$}_{\rm total}&=&\mbox{\boldmath$\omega$}_{L}+\mbox{\boldmath$\omega$}_{T}\nonumber\\ &=&-\frac{e}{m}\left(\frac{g-2}{2}+\frac{1}{\gamma}\right)B_z\widehat{{\mathbf k}},
\end{eqnarray}
in accordance with Thomas-BMT equation \cite{BMT} ( when the applied
magnetic field is only in the perpendicular direction). The
polarization of relativistic particles ($\gamma\rightarrow \infty$)
circulating in a transverse magnetic field precesses with a frequency
\begin{equation}
\left|\mbox{\boldmath$\omega$}_{\rm total}\right| =\frac{e B_z}{m}\left|\frac{g}{2}-1\right|,
\label{g1}
\end{equation}
which is used for measuring the anomalous magnetic moment of particles.

In the case of VSR the total precession turns out to be
\begin{eqnarray}
\mbox{\boldmath$\omega$}^{\rm VSR}_{\rm
  total}&=&\mbox{\boldmath$\omega$}_{L}+\mbox{\boldmath$\omega$}_{\rm
  VSR},
\end{eqnarray}
where $\mbox{\boldmath$\omega$}_{\rm VSR}$ has the form given in
Eq.~(\ref{vsr-freq}) and using Eq.~(\ref{accelr}) one gets
\begin{eqnarray}
\mbox{\boldmath$\omega$}_{\rm
  VSR}=-\frac{e}{m}\frac{\sqrt{\gamma^2-1}}{\gamma^2}B_z\widehat{{\mathbf
    i}}.
\end{eqnarray}
Therefore in VSR the total precession frequency of the particle will be
\begin{eqnarray}
\mbox{\boldmath$\omega$}^{\rm VSR}_{\rm
  total}=-\frac{e}{m}\frac{\sqrt{\gamma^2-1}}{\gamma^2}B_z\widehat{{\mathbf
    i}}-\frac{ge}{2m}B_z\widehat{{\mathbf k}},
\end{eqnarray}

So according to VSR theories particles circulating in a transverse
magnetic field will precess with a frequency which for high energy
particles is
\begin{equation}
\left|\mbox{\boldmath$\omega$}^{\rm VSR}_{\rm total}\right|=\left(\frac{e
  B_z}{m}\right) \frac{g}{2},
\label{g2}
\end{equation}
which is too large by a factor of $10^3$ compared to observations of
the precession rates of electrons and muons which is accurately
described by Eq.~(\ref{g1}).

%--------------------------------------------------------------------
\section{Conclusion}

We check that VSR theories reproduce the result of SR in case of
relativistic velocity addition. We show however that VSR theories fail
to reproduce one classic result of SR namely Thomas precession which
results in the spin orbit coupling interaction predicted by VSR
theories to be too large by a factor of 2 compared to observations of
the fine structure of atomic spectra. It is also interesting to note
that as there is no Lorentz violating parameter in VSR theories, it is
not possible to tune the parameter such that by doing so one can
obtain Thomas precession in these theories. It is the structure of the
proper subgroups of Lorentz group which leads to yield incorrect
Thomas precession in VSR theories.

Lorentz transformations have the property that two successive Lorentz
boosts is equivalent to a boost and a rotation, from the Lorentz
algebra $[K_i,K_j]=-\epsilon_{i j k} J_k$.  The algebra of $HOM(2)$ or
$SIM(2)$ is different so in the dynamics two successive VSR boosts
cannot be expressed as a combination of a VSR boost and VSR
rotation. Since accelerating observers have to be expressed in terms
of two separate boosts at $t$ and $t+\delta t$, the results for
accelerated observers differ between Lorentz transformation and VSR
transformations.

We also show that the equivalent of the BMT equation derived from VSR
theories results predicts the precession frequency of highly
relativistic particles in an external magnetic field to be too large
by a factor of $10^3$ compared to observations.

We conclude that although VSR can be used to derive many of the
classic results of Special Relativity it fails to give the correct
result for Thomas precession and is therefore ruled out as a
fundamental symmetry principle on which field theory of particles can
be constructed.

%------------------------------------------------------------------
\appendix
\section{Thomas precession and spin-orbital coupling in SR}
\label{thomas-SR}

In SR an instantaneous acceleration can be mimicked by a Lorentz
transformation combined with a rotation.  Consider an electron moving
in an orbit in $x-y$ plane around a nucleus. Let the velocity of the
electron in the rest frame of the nucleus be $\mathbf{v}=(v_x,0,0)$ at
some time $t$ and at a later time $t+\delta t$ the velocity be
$\mathbf{v}+\mathbf{\delta v}=(v_x+\delta v_x,\delta v_y,0)$. In SR
there is a Lorentz transformation which connects the instantaneous
electron velocity at time $t$ to its velocity $v_0=(1,0,0,0)$ in its
own rest frame,
\begin{equation}
\Lambda(\mathbf{v})v_0=v(t)
\label{l1}
\end{equation}
and similarly another Lorentz transformation connects the electron
velocity at time $t+\delta t$ with its velocity in its rest frame
$v_0$,
\begin{equation}
\Lambda(\mathbf{v}+\mathbf{\delta v}) v_0= v(t+\delta t).
\label{l2}
\end{equation}
The two velocities at different times can be connected with the
Lorentz transformation matrix $A^{\rm SR}_{T}$ ,
\begin{equation}
v(t+\delta t)=A^{\rm SR}_T v(t),
\end{equation}
which using Eq.~(\ref{l1}) and Eq.~(\ref{l2}) gives us
\begin{equation}
A^{\rm SR}_T=\Lambda(\mathbf{v}+\mathbf{\delta v})\Lambda^{-1}(\mathbf{v}).
\end{equation}
This matrix in the first order in $\mathbf{\delta v}$ gives us \cite{jackson},
\begin{eqnarray}
A_T^{\rm SR}=\left(
\begin{array}{ccccccc}
1 & & \gamma^2\delta v_x & & \gamma\delta v_y & & 0\\
\gamma^2\delta v_x & & 1 & & \frac{\gamma-1}{v_x}\delta v_y & & 0 \\
\gamma\delta v_y & & -\frac{\gamma-1}{v_x}\delta v_y & & 1 & & 0\\
0 & & 0 & & 0 & & 1
\end{array}
\right),
\end{eqnarray}
which can be written in terms of $J_i$ and $K_i$ as
\begin{eqnarray}
A_T^{\rm SR}%&=&I+\left( \gamma^2\delta v_x,\gamma\delta
%v_y,0\right)\cdot\mathbf{K}-\left(0,0,\frac{\gamma-1}{v_x}\delta v_y\right)\cdot\mathbf{J}\nonumber \\
=I-\Delta\mathbf{v}\cdot{\mathbf K}-\Delta{\mathbf \Omega}\cdot {\mathbf J},
\label{thomas_sr}
\end{eqnarray}
where $\Delta\mathbf{v}=-\left( \gamma^2\delta v_x,\gamma\delta
v_y,0\right)$ and $\Delta{\mathbf
\Omega}=\left(0,0,\frac{\gamma-1}{v_x}\delta v_y\right)$. To first
order in $\delta v_i$ the above equation can be written as
\begin{eqnarray}
A_T^{\rm SR}=A_{\rm boost}\left(\Delta\mathbf{v}\right)R\left(\Delta{\mathbf \Omega}\right)=R\left(\Delta{\mathbf \Omega}\right)A_{\rm boost}\left(\Delta\mathbf{v}\right),
\end{eqnarray}
where
\begin{eqnarray}
A_{\rm boost}\left(\Delta\mathbf{v}\right)&=&I-\Delta\mathbf{v}\cdot{\mathbf K},\\
R\left(\Delta{\mathbf \Omega}\right)&=&I-\Delta{\mathbf \Omega}\cdot {\mathbf J},
\end{eqnarray}
and the rotation angle
\begin{equation}
{\Delta \mathbf{\Omega}}=\left(0,0, \frac{\gamma-1}{v_x} \delta v_y\right).
\end{equation}
So the electron rotates with respect to the frame of the nucleus with
an angular velocity
\begin{equation}
\mbox{\boldmath$\omega$}_{T}=-\frac{{\Delta \mathbf{\Omega}}}{\delta t}=\left(0,0, -\frac{\gamma^2}{\gamma+1} v_x a_y\right)\simeq \left(0,0, -\frac{1}{2} v_x a_y\right),
\label{thomas-freq}
\end{equation}
where the last equality is obtained assuming non-relativistic limits.
The spin of the electron precesses in the rest frame of the nucleus as
\begin{equation}
\left.\frac{d {\bf s}}{dt}\right|_{\rm nucleus-frame}= \left.\frac{d {\bf s}}{dt}\right|_{\rm e-frame} +
\mbox{\boldmath$\omega$}_{T}\times {\bf s}.
\label{ds-dt}
\end{equation}
This extra precession known as the Thomas precession corresponds to an
interaction
\begin{equation}
U_T=\mbox{\boldmath$\omega$}_{T}\cdot {\bf s}.
\end{equation}
The electron's magnetic moment has the interaction energy
\begin{equation}
U=-\mbox{\boldmath$\mu$} \cdot \mathbf{B^\prime},
\label{sob}
\end{equation}
where $\mbox{\boldmath$\mu$}=\frac{g e}{2 m} \mathbf{s}$ is the
magnetic moment of the electron and $\mathbf{B^\prime}=
-\mathbf{v}\times\mathbf{ E}$ is the effective magnetic field in the
rest frame of the electron and ${\mathbf E} $ is the electric field of
the nucleus given as
\begin{equation}
\mathbf{E}= -\frac{\bf r}{e r } \frac{dV}{dr}.
\end{equation}
The total spin-orbital interaction energy of the electron is therefore
\begin{equation}
H_{SO}=-\mbox{\boldmath$\mu$} \cdot
\mathbf{B^\prime}+\mbox{\boldmath$\omega$}_{T}\cdot {\bf s}.
\end{equation}
Using the fact that $a_y= e E_y/m$, the spin-orbital energy turns out to be
\begin{equation}
H_{SO}={\frac{g-1}{2 m^2} } {\bf s}\cdot {\bf L}\frac{1}{r}
\frac{dV}{dr}.
\label{hsosr}
\end{equation}
The measurement of the Zeeman-splitting of spectral lines in a
magnetic field shows that the gyromagnetic ratio of the electron is $g
\simeq 2$. If the Thomas precession was absent the spin-orbital
coupling term would have a factor of $g$ in the first bracket instead
of $(g-1)$ which would have resulted in a factor of 2 discrepancy with
observations.

%%%%%%%%%%%%%%%%%%%%%%%%%%%%%%%%%%%%%%%%%%%%%%
%\section*{References}

\end{document}